\documentclass[longauth]{aa} 

\usepackage{graphicx}
\usepackage{txfonts}

\def\ltsima{$\; \buildrel < \over \sim \;$}
\def\simlt{\lower.5ex\hbox{\ltsima}}
\def\gtsima{$\; \buildrel > \over \sim \;$}
\def\simgt{\lower.5ex\hbox{\gtsima}}

\def\cm2{\mbox{$\mbox{cm}^{-2}$}}
\def\cm3{\mbox{$\mbox{cm}^{-3}$}}
\def\h2{\mbox{$_{\mbox{\tiny H2}}$}}
\begin{document} 

\title{Characterizing filaments in regions of high-mass star formation: 
High-resolution submilimeter imaging of the massive star-forming complex NGC 6334 with ArT\'eMiS\thanks{This publication is based on data acquired with the Atacama Pathfinder Experiment (APEX) in ESO program 091.C-0870. APEX is a collaboration between the Max-Planck-Institut f\"ur Radioastronomie, the European Southern Observatory, and the Onsala Space Observatory.} }

\titlerunning{High-resolution submm imaging of NGC 6334 with ArT\'eMiS}
          
    \author{Ph.~Andr\'e\inst{1}
          \and
          V.~Rev\'eret\inst{1}
          \and
          V.~K\"onyves\inst{1}
          \and
          D.~Arzoumanian\inst{1}
          \and 
          J.~Tig\'e\inst{2}
          \and
          P.~Gallais\inst{1}
                    \and
          H.~Roussel\inst{3}
          \and
	 J.~Le~Pennec\inst{1}
          \and
          L.~Rodriguez\inst{1}
          \and
          E.~Doumayrou\inst{1}
          \and
          D.~Dubreuil\inst{1}
          \and
          M.~Lortholary\inst{1}
          \and
          J.~Martignac\inst{1}
	\and
          M.~Talvard\inst{1}
                    \and
          C.~Delisle\inst{1}        
          \and
          F.~Visticot\inst{1}
          \and
          L.~Dumaye\inst{1}                              
          \and 
          C.~De~Breuck\inst{4}
          \and
          Y.~Shimajiri\inst{1}
	 \and 
          F.~Motte\inst{1}
           \and 
          S.~Bontemps\inst{5}
           \and
          M.~Hennemann\inst{1}
	 \and 
          A.~Zavagno\inst{2}
	 \and 
          D.~Russeil\inst{2}
          \and 
          N.~Schneider\inst{5,6}
	 \and 
          P.~Palmeirim\inst{2,1}
	 \and 
          N.~Peretto\inst{7}                                
 	\and
         T.~Hill\inst{1}
 	\and
         V.~Minier\inst{1}
         \and
         A.~Roy\inst{1}
         \and 
         K.L.J.~Rygl\inst{8}                        
          }

\authorrunning{Ph.~Andr\'e, V.~Rev\'eret, V.~K\"onyves et al.}

   \institute{Laboratoire AIM, CEA/DRF--CNRS--Universit\'e Paris Diderot, IRFU/Service d'Astrophysique, C.E. Saclay,
              Orme des Merisiers, 91191 Gif-sur-Yvette, France\\
              \email{pandre@cea.fr}
          \and Aix Marseille Universit\'e, CNRS, LAM (Lab. d'Astrophysique de Marseille), UMR 7326, 13388 Marseille, France    
 	 \and Institut d'Astrophysique de Paris, Sorbonne Universit\'es,
      UPMC Univ. Paris 06, CNRS, UMR 7095, 75014 Paris, France
 	 \and
	   European Southern Observatory,  Karl Schwarzschild Str. 2, 85748 Garching bei Munchen, Germany
	 \and Univ. Bordeaux, LAB, UMR 5804, 33270, Floirac, France	
	 \and I. Physik. Institut, University of Cologne, 50937 K\"oln, Germany
	 \and School of Physics \& Astronomy, Cardiff University, The Parade, Cardiff CF24 3AA, UK
	 \and INAF-Osservatorio di Radioastronomia, Via P. Gobetti 101, 40129 Bologna, Italy
             }
          
   \date{Received 25 February 2016 ; accepted 18 May 2016}

 
\abstract
{
{\it Herschel} observations of nearby molecular clouds 
suggest that interstellar filaments and prestellar cores represent two fundamental steps in the star formation process.
The observations support a picture of low-mass star formation according to which $\sim 0.1$ pc-wide filaments
form first in the cold interstellar medium, probably as a result of large-scale compression of interstellar matter 
by supersonic turbulent flows, and then prestellar cores arise from gravitational fragmentation of the densest filaments. 
Whether this scenario also applies to regions of high-mass star formation is an open question, 
in part because the resolution of {\it Herschel} is insufficient to resolve the inner width of filaments in the nearest 
regions of massive star formation. 
}
{
In an effort to characterize the inner width of filaments in high-mass star forming regions, we imaged 
the central part of the NGC6334 complex at a factor of $> 3$ higher resolution than {\it Herschel} at 350~$\mu$m.
}
{
We used the large-format bolometer camera ArT\'eMiS on the APEX telescope 
and combined the high-resolution ArT\'eMiS data at 350~$\mu$m with {\it Herschel}/HOBYS data at 70--500~$\mu$m
to ensure good sensitivity to a broad range of spatial scales. 
This allowed us 
to study the structure of the main narrow filament of the complex with a resolution 
of $8\arcsec$ or $< 0.07$~pc at $d \sim 1.7$~kpc. 
}
{
Our study confirms that this filament is a very dense, massive linear structure 
with a line mass ranging from $\sim 500\, M_\odot$/pc to $\sim 2000\, M_\odot$/pc over nearly $10\,$pc.  
It also demonstrates for the first time that its inner width remains as narrow as $W \sim 0.15\pm 0.05\,$pc all along 
the filament length, within a factor of $< 2$ of the characteristic $0.1\,$pc value found with {\it Herschel} for lower-mass filaments 
in the Gould Belt. 
}
{
While it is not completely clear whether the NGC~6334 filament will form massive stars or not in the future, 
it is two to three orders of magnitude denser than the majority of filaments observed in Gould Belt clouds, 
and yet has a very similar inner width. This points to a common physical mechanism for setting the filament width 
and suggests that some important structural properties of nearby clouds also hold in high-mass star forming regions. 
}


   \maketitle
%


\section{Introduction}

Understanding star formation 
is a fundamental issue in modern astrophysics (e.g., \citealp{McKeeOstriker2007}).
Very significant observational progress has been made 
on this topic thanks to far-infrared and submillimeter imaging surveys with 
the {\it Herschel} Space Observatory. 
In particular, the results from the {\it Herschel} ``Gould Belt'' survey (HGBS) 
confirm the omnipresence of  filaments in nearby clouds 
and suggest an intimate connection between the filamentary structure 
of the interstellar medium (ISM) 
and the formation process of low-mass prestellar cores \citep[][]{Andre+2010}. 
While molecular clouds were already known to exhibit large-scale filamentary 
structures for quite some time 
\citep[e.g.][and references therein]{SchneiderElmegreen1979, Myers2009}, 
{\it Herschel} observations now demonstrate 
that these filaments are truly ubiquitous in the cold ISM \citep[e.g.][]{Molinari+2010,Henning+2010,Hill+2011}, 
probably make up a dominant fraction of the dense gas in molecular clouds \citep[e.g.][]{Schisano+2014,Konyves+2015}, 
and present a high degree of universality in their properties \citep[e.g.][]{Arzoumanian+2011}. 
Therefore, interstellar filaments likely play a central role in the star formation process \citep[e.g.][]{Andre+2014}. 
A detailed analysis of their radial column density profiles 
shows that, at least in the nearby clouds of the Gould Belt, 
filaments are characterized 
by a very narrow distribution of 
inner widths $W$ with a typical FWHM 
value $\sim 0.1\,$pc (much larger than the $\sim 0.01\,$pc resolution provided by {\it Herschel} at  the distance $\sim 140\,$pc 
of the nearest clouds)  and a dispersion of less than a factor of 2 
\citep[][]{Arzoumanian+2011,KochRosolowsky2015}.
The origin of this common inner width of interstellar filaments 
is not yet well understood.
A possible interpretation  
is that it corresponds to the sonic scale below which interstellar turbulence becomes subsonic in diffuse, non-star-forming molecular gas 
\citep[cf.][]{Padoan+2001,Federrath2016}. 
Alternatively, this characteristic inner width of filaments may be set by the dissipation mechanism 
of magneto-hydrodynamic (MHD) waves 
\citep[e.g.][]{HennebelleAndre2013}. 
A possible manifestation of such MHD waves may actually have been found   
in the form of braided velocity-coherent substructure in the case of the Taurus B211--3 filament \citep{Hacar+2013}.
Another major result from {\it Herschel} in nearby clouds 
is that most ($> 75\% $) low-mass 
prestellar cores and protostars are found in dense,  ``supercritical'' filaments for which 
the mass per unit length  $M_{\rm line}$ exceeds the critical line mass of nearly isothermal, long cylinders  
\citep[e.g.][]{InutsukaMiyama1997}, 
$M_{\rm line, crit} = 2\, c_s^2/G \sim 16\, M_\odot$/pc,  
where $c_{\rm s} \sim 0.2$~km/s is the isothermal sound speed for molecular gas at $T \sim 10$~K 
\citep[e.g.][]{Konyves+2015}.
These {\it Herschel} findings support a scenario for low-mass star formation 
in two main steps \citep[cf.][]{Andre+2014}: 
First, large-scale compression of interstellar material in supersonic MHD flows 
generates a cobweb of $\sim 0.1$ pc-wide filaments in the ISM; 
second, the densest filaments fragment into 
prestellar cores (and subsequently protostars) by gravitational instability above $M_{\rm line, crit} $, 
while simultaneously growing in mass through accretion of background cloud material.

In addition to the relatively modest 
filaments found in non star forming and low-mass star forming 
clouds, where $M_{\rm line}$ rarely exceeds ten times the thermal value of $M_{\rm line, crit} $, 
significantly denser and more massive filamentary structures have also been observed in 
the most active giant molecular clouds (GMCs) of the Galaxy, 
and may be the progenitors of young massive star clusters. 
The DR21 main filament or ``ridge'' is probably the most emblematic case of such a massive elongated structure 
with about $20000\, M_\odot $ inside a $4.5\,$pc long structure (i.e.,  $M_{\rm line} \sim 4500\, M_\odot $/pc) 
\citep{Motte+2007, Schneider+2010, Hennemann+2012}.
Other well-known ridges include Orion$\,$A \citep{Hartmann+2007}, 
Vela-C  \citep{Hill+2011,Hill+2012}, IRDC G035.39--00.33 \citep{Nguyen+2011}, 
and W43-MM1 \citep{Nguyen+2013,Louvet+2014}. 
These ridges, which exceed the critical line mass of an isothermal filament by up to two orders of magnitude, 
are believed to be in a state of global collapse, 
to be fed by very high accretion rates on large scales 
\citep{Schneider+2010,Peretto+2007,Peretto+2013}, 
and to continuously form stars and clusters. 
The formation of these ridges is not yet well understood but may result 
from the large-scale collapse of a significant portion of a GMC 
\citep{Hartmann+2007, Schneider+2010}.

Whether the low-mass star formation scenario summarized above 
-- or an extension of it -- also applies to regions  
dominated by hyper-massive clumps and ridge-like structures 
\citep{Motte+2016} 
is not yet known. 
In particular, further work is needed to confirm that the inner width of interstellar filaments remains close to $\sim 0.1$~pc
in regions of massive star formation 
beyond the Gould Belt, where the moderate angular resolution of  {\it Herschel} (HPBW $\sim \, $18--36\arcsec 
~at $\lambda = \, $250--500~$\mu$m) is insufficient to resolve this characteristic scale.

At a distance of $\sim\, $1.7~kpc, 
NGC~6334 is a very active complex of massive star formation \citep[][]{PersiTapia2008,Russeil+2013}
with about 150 associated luminous stars of O- to B3-type \citep[][]{Neckel1978,Bica+2003,Feigelson+2009}. 
At far-infrared and (sub)millimeter wavelengths, the central part of NGC\,6334 consists of a 10\,pc-long elongated
structure including two major high-mass star-forming clumps and a narrow filament \citep[e.g.][]{Sandell2000, Tige+2016}. 
The filament is particularly prominent in ground-based (sub)millimeter continuum images 
where extended emission is effectively filtered \citep[e.g.][]{Munoz+2007, Matthews+2008}. 
It apparently forms only low-mass stars \citep{Tige+2016}, except perhaps at its end points, 
in marked contrast with the high-mass clumps which host several protostellar ``massive dense cores'' 
\citep{Sandell2000,Tige+2016}.
The multi-wavelength coverage and high dynamic range of {\it Herschel} observations from the HOBYS key project \citep{Motte+2010}
gave an unprecedented view of the column density and dust temperature structure of NGC~6334 
with a resolution limited to $36\arcsec $ or $0.3\, $pc when the 500\,$\mu$m band was used 
(\citealp{Russeil+2013} and \citealp{Tige+2016}). 
The NGC~6334 filament 
has a line mass 
approaching $M_{\rm line}  \sim 1000\, M_\odot$/pc 
and features column densities close to or above 10$^{23}$\,cm$^{-2}$ over about 10\,pc 
along its length  
\citep[e.g.][]{Matthews+2008,Zernickel+2013}. 

Here, we report the results of high-resolution ($8\arcsec $) 350~$\mu$m dust continuum mapping observations of the central part of NGC~6334 
with the ArT\'eMiS bolometer camera on the APEX 12-m telescope. 
The $\sim 8\arcsec $ resolution of ArT\'eMiS at 350~$\mu$m, corresponding to $\sim 0.068$~pc at the distance of NGC~6334, has 
allowed us to resolve, for the first time, the transverse size of the main 
filament in this complex. 
Section~\ref{obs_set} describes the instrument and provides details about the observing run and data reduction. 
Section~\ref{obs_ana} presents our mapping results, which are discussed in Section~\ref{dis}.


\section{ArT\'eMiS observations and data reduction}
\label{obs_set}

Our 350~$\mu$m observations of NGC~6334 were obtained in July--September 2013 and June 2014 with the 
ArT\'eMiS\footnote{See http://www.apex-telescope.org/instruments/pi/artemis/\\ 
ArT\'eMiS stands for ``ARchitectures de bolom\`etres pour des TElescopes \`a grand champ de vue dans le domaine sub-MIllim\'etrique au Sol'' in French.}   
camera on the Atacama Pathfinder Experiment (APEX) telescope located at an altitude of 5100~m at Llano de Chajnantor in Chile. 
ArT\'eMiS is a large-format bolometer array  camera, built by CEA/Saclay and installed in the Cassegrain cabin of APEX, which will eventually have 
a total of 4608 
pixels observing at 350~$\mu$m and 450~$\mu$m 
simultaneously \citep[][]{Talvard+2010, Reveret+2014}.

ArT\'eMiS employs the technology successfully developed by CEA for the PACS photometer instrument in the 60--210~$\mu$m wavelength regime 
on the {\it Herschel} Space Observatory \citep[e.g.][]{Billot+2006}. 
Unlike the 
LABOCA camera on APEX, the ArT\'eMiS instrument does not use feedhorns to concentrate the incoming submillimeter radiation, 
but planar bare arrays of $16\times 18$ silicon bolometer pixels each which act like a CCD camera does in the optical domain. 
The 2013 and 2014 incarnations of ArT\'eMiS used for these observations  
were equipped with a 350~$\mu$m focal plane of four and eight such sub-arrays of $16\times 18$ pixels, respectively. 
The number of working pixels was about 1050 in 2013 and 1650 in 2014. 
The instantaneous field of view of the camera was $\sim 2.1\arcmin \times 2.4\arcmin $ in 2013 and $\sim 4.3\arcmin \times 2.4\arcmin $ in 2014, 
and was essentially fully sampled.
ArT\'eMiS features a closed-cycle cryogenic system built around a pulse tube cooler (40~K and 4~K) 
coupled to a double stage helium sorption cooler ($\sim 300$~mK). During the 2013 and 2014 observing campaigns, 
the typical hold time of the cryostat at 260~mK between two remote recycling procedures at the telescope was $> 48$ hours. 

\begin{figure*}[!htp]
      \centering
  \resizebox{9.1cm}{!}{
 \includegraphics[angle=0]{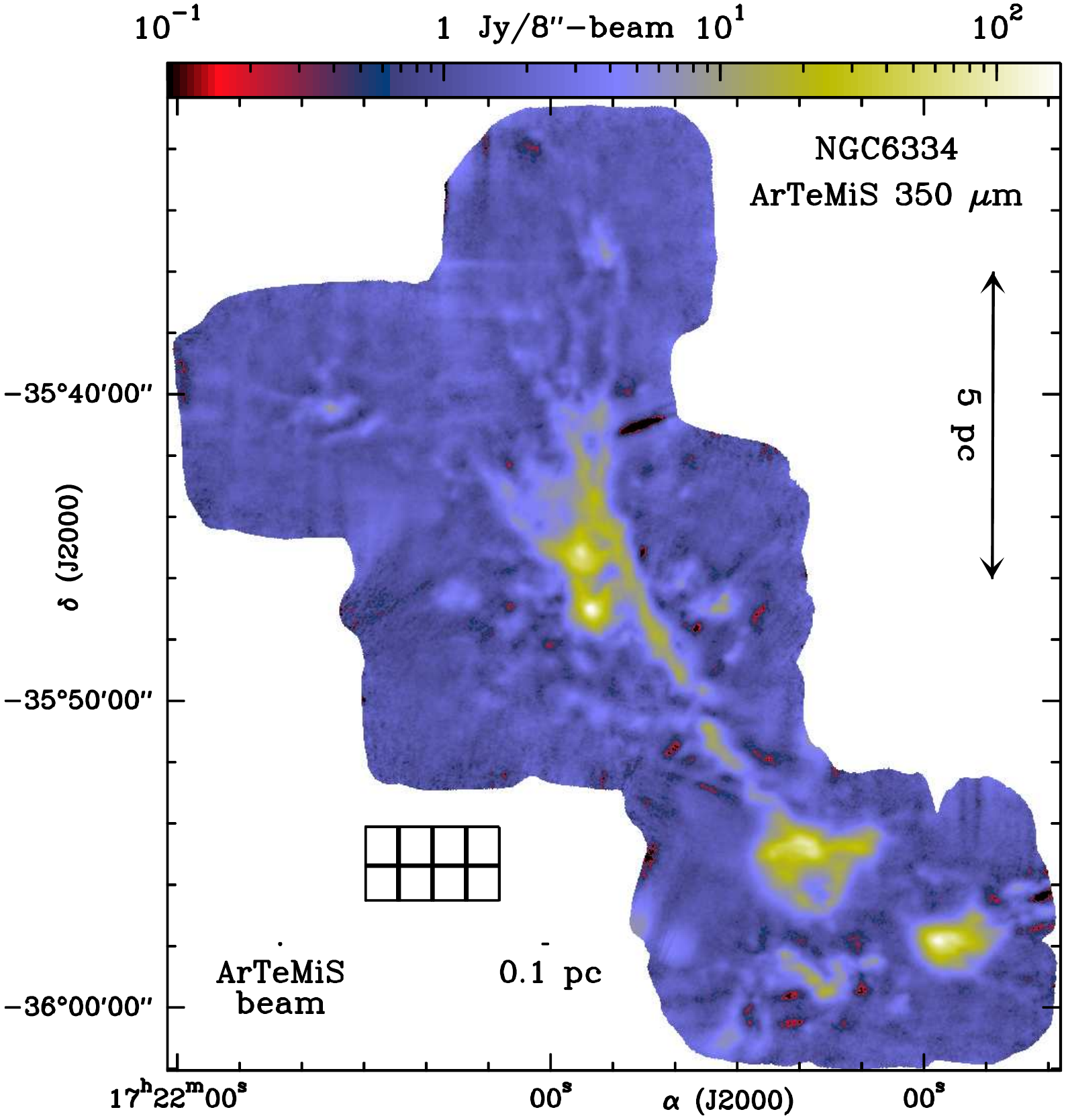}} 
       \hspace{0.05mm}
   \resizebox{9.1cm}{!}{      
\includegraphics[angle=0]{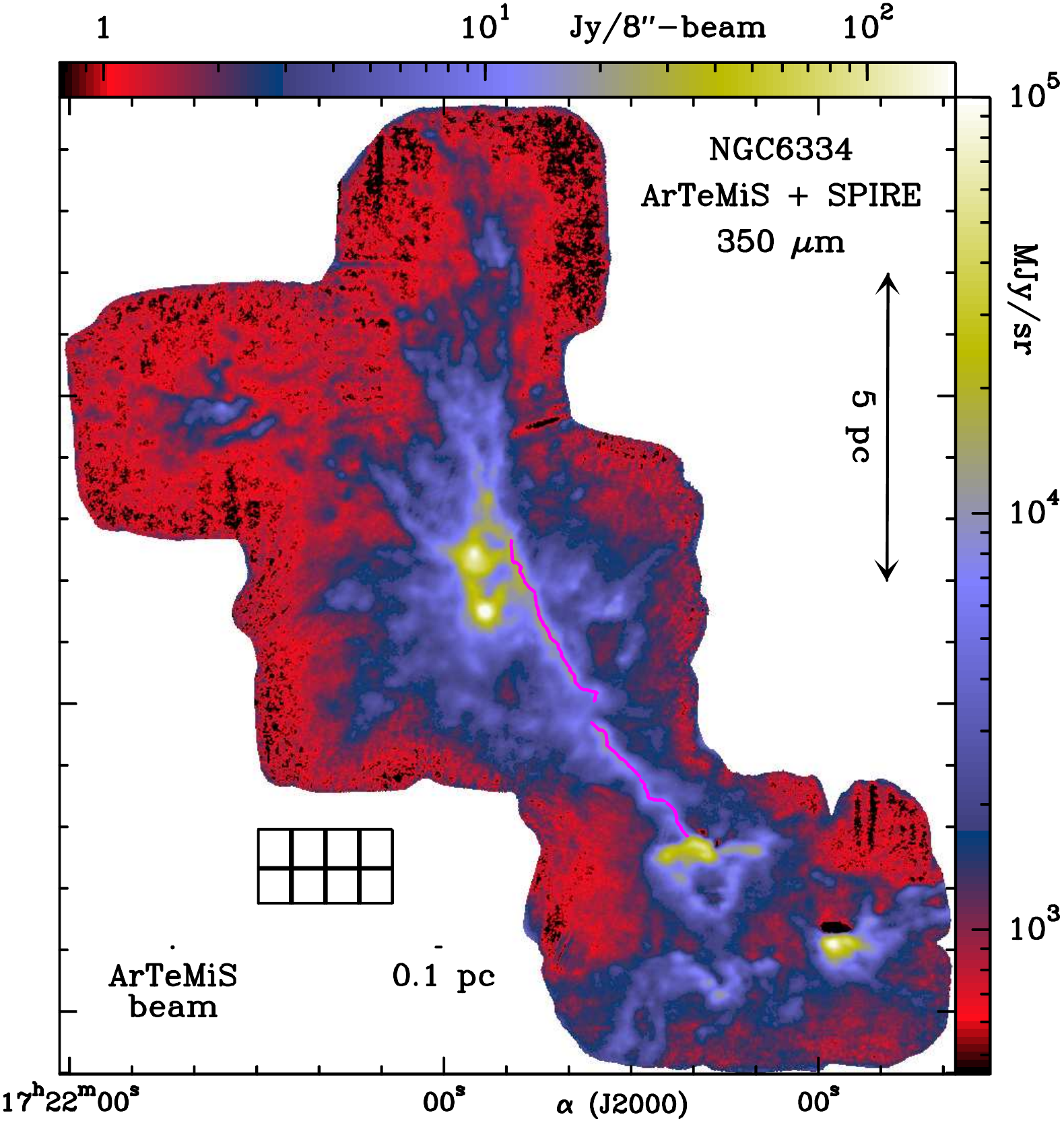}} 
   \caption{
   {\bf(a)} ArT\'eMiS 350 $\mu$m dust continuum mosaic of the central part of the NGC~6334 
star-forming complex.  The effective half-power beam width (HPBW) resolution is $8\arcsec$ (or $ 0.07$~pc at $d=\, $1.7 kpc), 
a factor of $> 3$ 
higher than the {\it Herschel}/HOBYS SPIRE 350~$\mu$m map \citep{Russeil+2013,Tige+2016}. 
   {\bf(b)} High dynamic range 350 $\mu$m dust continuum map of the same field obtained by 
combining the ArT\'eMiS high-resolution data shown in the left panel with the {\it Herschel}/SPIRE 350 $\mu$m data 
providing better sensitivity to large-scale emission. The effective angular resolution is also $8\arcsec $ (HPBW). 
The crest of the 
main 
filament as traced by the DisPerSE algorithm \citep[][]{Sousbie2011} 
is marked by the magenta curve. 
The footprint of the $4\times 2$ sub-arrays of the ArT\'eMiS camera indicating the instantaneous field of view on the sky 
is shown in both panels. (The instantaneous field of view for the data taken in 2013 corresponds to the central $2\times 2$ sub-arrays.)
 }
              \label{artemis_maps}
    \end{figure*}

A total of 35 individual maps, corresponding to a total telescope time of $\sim 13$~hr (excluding pointing, focusing, and calibration scans),  
were obtained with ArT\'eMiS at $350\, \mu$m toward the NGC~6334 region using a total-power, on-the-fly scanning mode. 
Each of these maps consisted in a series of scans taken either in Azimuth or at a fixed angle with respect to the Right Ascension axis. 
The scanning speed ranged from 20\arcsec /sec to 30\arcsec /sec and 
the cross-scan step between consecutive scans  
from $3\arcsec $ to $10\arcsec $. 
The sizes of the maps ranged from $3.5\arcmin \times 3.5\arcmin $ to $11.5\arcmin \times 10\arcmin $. 
The atmospheric opacity at zenith was measured using skydips with ArT\'eMiS and was found to vary between 0.45 and 1.85 at $\lambda = 350\, \mu$m.
This is equivalent to an amount of precipitable water vapor (PWV) from $\sim 0.25$~mm to $\sim 0.9$~mm with a median value of 0.53~mm. 
The median elevation of NGC~6334 was $\sim 58\degr $ corresponding to a median airmass of 1.18.

A dedicated pointing model was derived for ArT\'eMiS after the first days of 
commissioning observations in July 2013 
and was found to yield 
good results ($3\arcsec $ overall rms error) throughout  the ArT\'eMiS observing campaign.  
Absolute calibration was achieved by taking both short `spiral' scans and longer on-the-fly beam maps of the primary calibrators Mars and Uranus.  
During the mapping of NGC~6334, regular pointing, focus, and calibration checks were made
by observing `spiral' scans of the nearby secondary calibrators G5.89, G10.47, G10.62, and IRAS~16293. 
The maximum deviation observed between two consecutive pointing checks was $\sim 3\arcsec $. 
The absolute pointing accuracy is estimated to be $\sim 3\arcsec $ and the absolute calibration uncertainty to be $\sim 30\% $.  

The median value of the noise equivalent flux density (NEFD) per detector was $\sim 600$~mJy$\,$s$^{1/2}$, 
with best pixel values at $\sim 300$~mJy$\,$s$^{1/2}$.
The pixel separation between detectors on the sky was $\sim 3.9\arcsec $, corresponding to Nyquist spacing at $ 350\, \mu$m.
As estimated from our maps of Mars,  
the main beam had a full width at half maximum (FWHM) of $ 8.0 \pm 0.1\arcsec $ and contained $\sim 70\% $ of the power, the 
rest being distributed in an ``error beam'' extending up to an angular radius of $\sim 40\arcsec $ (see blue solid curve
in Fig.~\ref{filament_profile}a in Sect.~\ref{obs_ana} below for the beam profile). 

Online data reduction at the telescope was performed with the BoA software 
developed for LABOCA \citep[][]{Schuller2012}. 
Offline data reduction, including baseline subtraction, removal of correlated skynoise and $1/f$ noise, and  
subtraction of uncorrelated $1/f$ noise was performed with in-house IDL routines, including tailored versions
of the Scanamorphos software routines 
which exploit the redundancy in the mapping raw data, especially data taken with filled arrays. 
The Scanamorphos algorithm, as developed to process {\it Herschel} observations, is described in
depth in \citet{Roussel2013}.
To account for the specificities of the observations discussed here, it had to
be modified.
The destriping step for long scans had to be deactivated, as well as the
average drift subtraction in scans entirely filled with sources, and 
a sophisticated filter had to be applied to subtract the correlated skynoise. 
This filter involves a comparison between the signal of all sub-arrays
at each time, and a protection of compact sources by means of a mask initialized
automatically, and checked manually.

\begin{figure*}[!htp]
      \centering
  \resizebox{9.8cm}{!}{
 \includegraphics[angle=0]{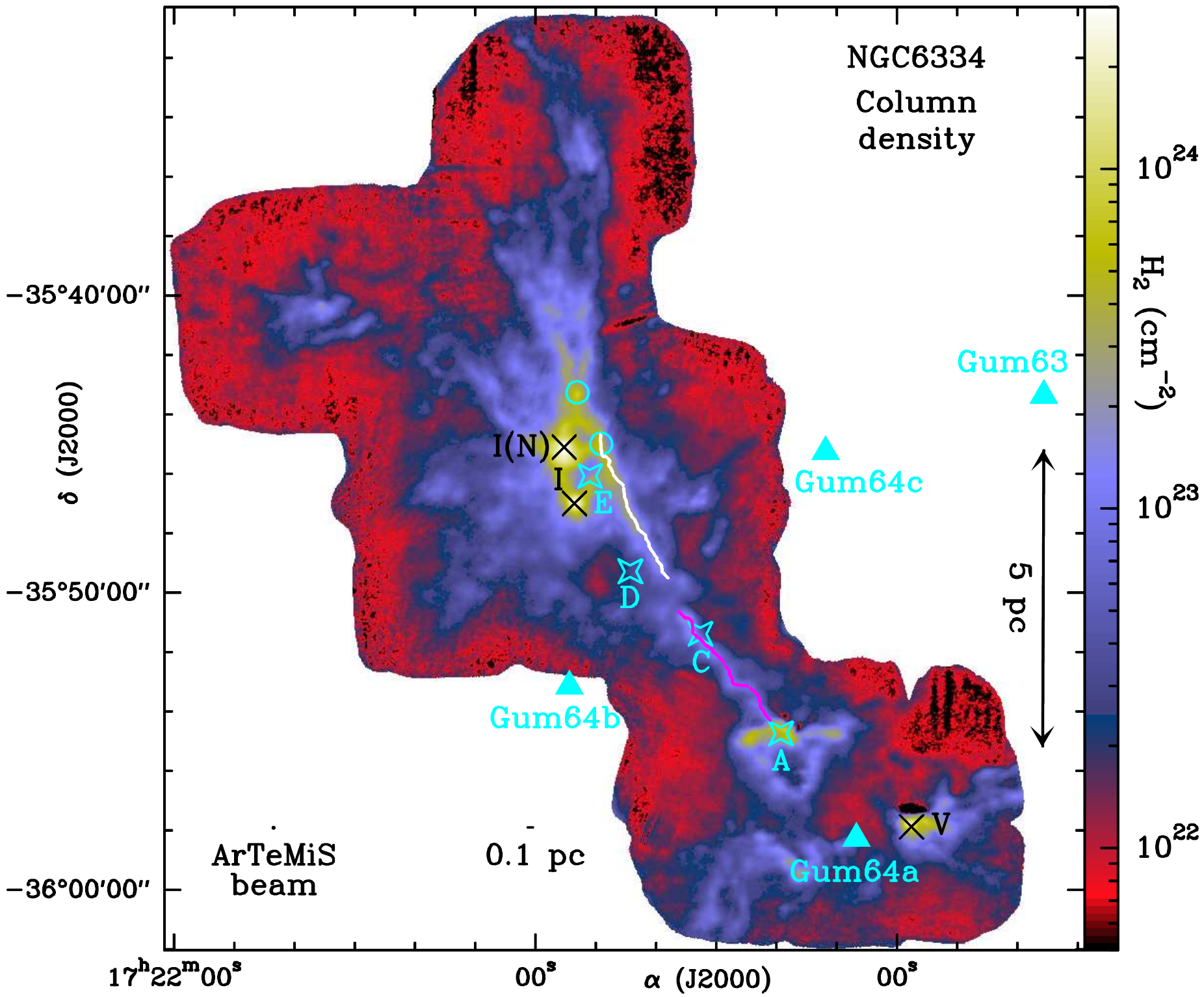}} 
       \hspace{1mm}
   \resizebox{8.1cm}{!}{
\includegraphics[angle=0]{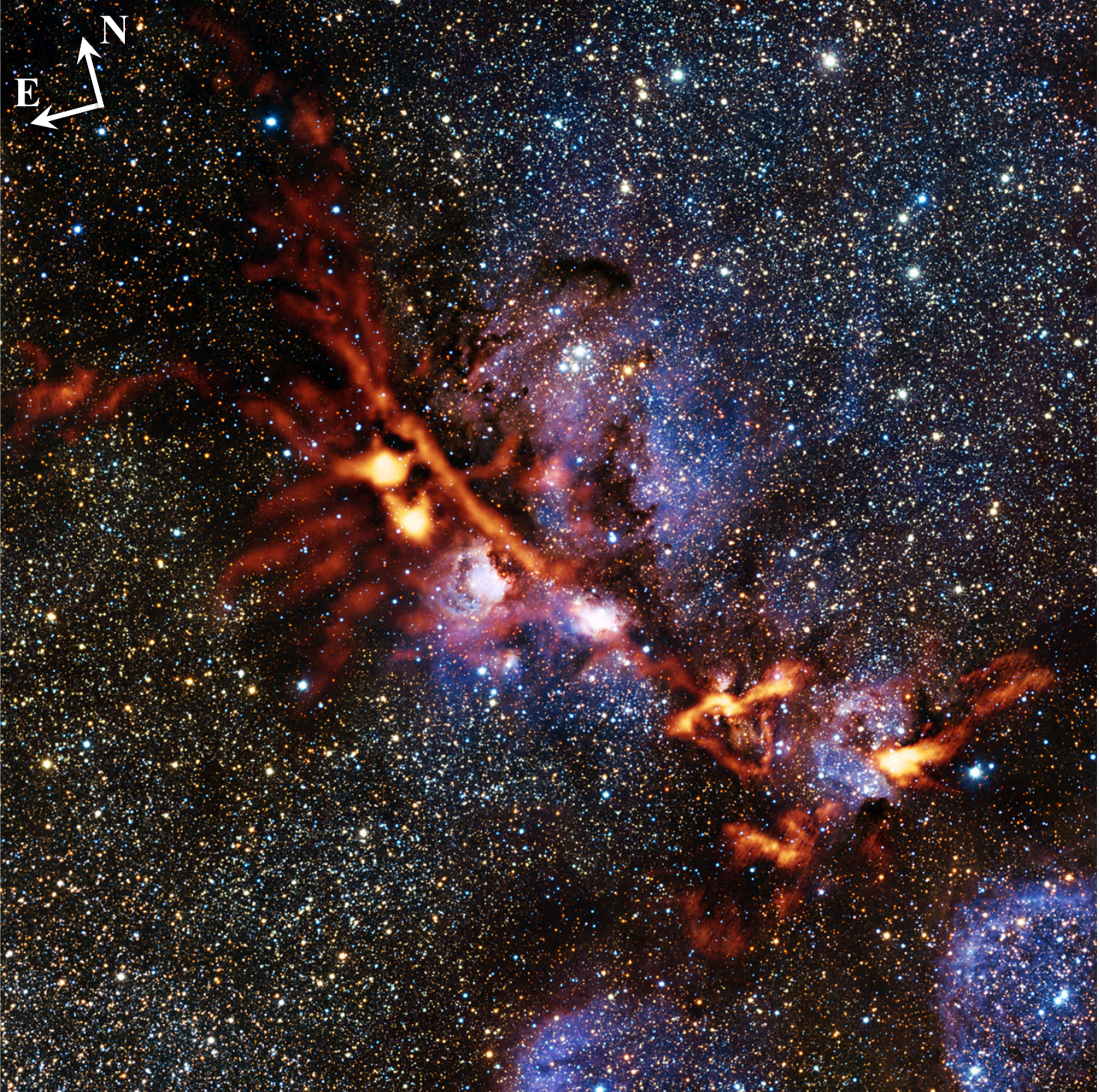}} 
   \caption{
   {\bf(a)} Approximate H$_2$ column density map of the central part of the NGC~6334 star-forming complex derived 
from the combined ArT\'eMiS $+$ SPIRE 350 $\mu$m image shown in Fig.~\ref{artemis_maps}b, assuming optically thin dust 
emission at the temperature given by the {\it Herschel}/HOBYS dust temperature map of \citet{Tige+2016} (see also  
\citealp{Russeil+2013}). 
The effective angular resolution is $8\arcsec $ (HPBW). 
The crest of the northern part of the NGC~6334 main filament as traced by the DisPerSE algorithm \citep[][]{Sousbie2011} 
is marked by the white curve, while the crest of the southern part is shown by the magenta curve.  
Black crosses and roman numerals (I, V) denote bright far-infrared sources \citep[][]{Kraemer+1999}. 
Cyan open circles mark two candidate starless ``massive dense cores'' from  \citet{Tige+2016}. 
Cyan open diamonds and alphabetical letters A--E indicate compact HII regions detected in the $6\,$cm radio continuum with the VLA 
\citep{Rodriguez+1982}, and cyan filled triangles mark diffuse HII regions traced as diffuse H$\alpha$ emission 
nebulosities \citep{Gum1955}. 
   {\bf(b)}~ArT\'eMiS 350 $\mu$m image (orange) overlaid on a view of the same region taken at near-infrared wavelengths 
   with the ESO VISTA telescope (see ESO photo release http://www.eso.org/public/images/eso1341a/ -- Credit: ArT\'eMiS team/ESO/J. Emerson/VISTA). 
 }
              \label{artemis_add_maps}
\end{figure*}

\section{Mapping results and radial profile analysis}
\label{obs_ana}

By co-adding the 35 individual ArT\'eMiS maps of NGC~6334, we obtained the $350\, \mu$m mosaic shown in 
Fig.~\ref{artemis_maps}a. 
The typical rms noise in this mosaic is $\sigma \sim 0.2\,$Jy/8\arcsec -beam. 
As usual with total-power ground-based submillimeter continuum observations, the ArT\'eMiS raw data were 
affected by a fairly high level of skynoise, strongly correlated over the multiple detectors of the focal plane. 
Because of the need to subtract this correlated skynoise to produce a meaningful image, the mosaic of Fig.~\ref{artemis_maps}a 
is not sensitive to angular scales larger than the instantaneous field of view of the camera $\sim 2\arcmin $. 
The large-scale background intensity (e.g. zero level) in the image of Fig.~\ref{artemis_maps}a is therefore not constrained 
by the ArT\'eMiS observations and has been arbitrarily set to a small positive value 
(corresponding approximately to $\sim 5\sigma $) 
to facilitate the display using a logarithmic intensity scale. 
To restore the missing large-scale information, we combined the ArT\'eMiS data with the SPIRE $350\, \mu$m data 
from the {\it Herschel} HOBYS key project \citep[][]{Motte+2010,Russeil+2013} 
employing a technique similar to that used in combining millimeter interferometer observations with single-dish data. 
In practice, this combination was achieved with the task ``immerge'' in the Miriad software package \citep[][]{Sault+1995}. 
Immerge combines two datasets in the Fourier domain after determining an optimum calibration factor to align the flux 
scales of the two input images in a common annulus of the uv plane. Here, a calibration factor of 0.75 had to be applied 
to the original ArT\'eMiS image to match the flux scale of the SPIRE $350\, \mu$m image over a range of baselines from 
0.6~m (the baseline b sensitive to angular scales ${\rm b}/\lambda \sim 2\arcmin $ at $350\, \mu$m)  
to 3.5~m (the diameter of the {\it Herschel} telescope). 
The magnitude of this factor is consistent with the absolute 
calibration uncertainty of $\sim 30\% $ quoted in Sect.~\ref{obs_set}.
The resulting combined $350\, \mu$m image of NGC~6334 has an effective resolution of $\sim 8\arcsec $ (FWHM) 
and is displayed in Fig.~\ref{artemis_maps}b.

To determine the location of the crest of the main 
filament in NGC~6334, we applied the DisPerSE algorithm \citep[][]{Sousbie2011}  
to the combined $350\, \mu$m image. 
The portion of the filament analyzed in some detail below was selected so as to avoid 
the confusing effects of massive young stars and protostellar ``massive dense cores'' (MDCs) \citep[cf.][]{Tige+2016}. 
It nevertheless includes one candidate starless MDC at its northern end (see Fig.~\ref{artemis_add_maps}a). 
The corresponding crest 
is shown as a magenta solid curve in Fig.~\ref{artemis_maps}b. 

\begin{figure*}[!htp]
      \centering
    \resizebox{9.75cm}{!}{      
\includegraphics[angle=0]{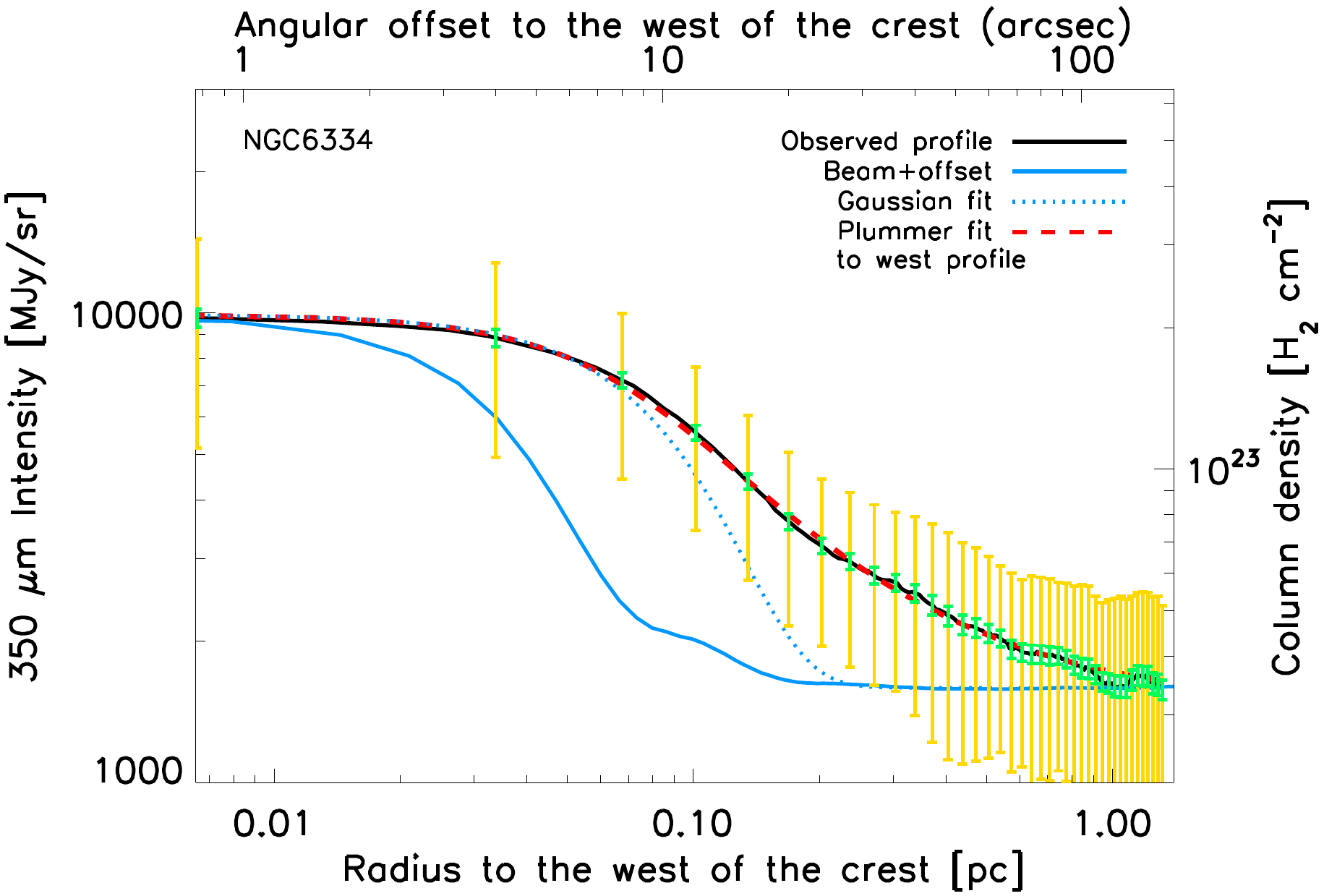}}
       \hspace{1mm}
  \resizebox{8.cm}{!}{
 \includegraphics[angle=0]{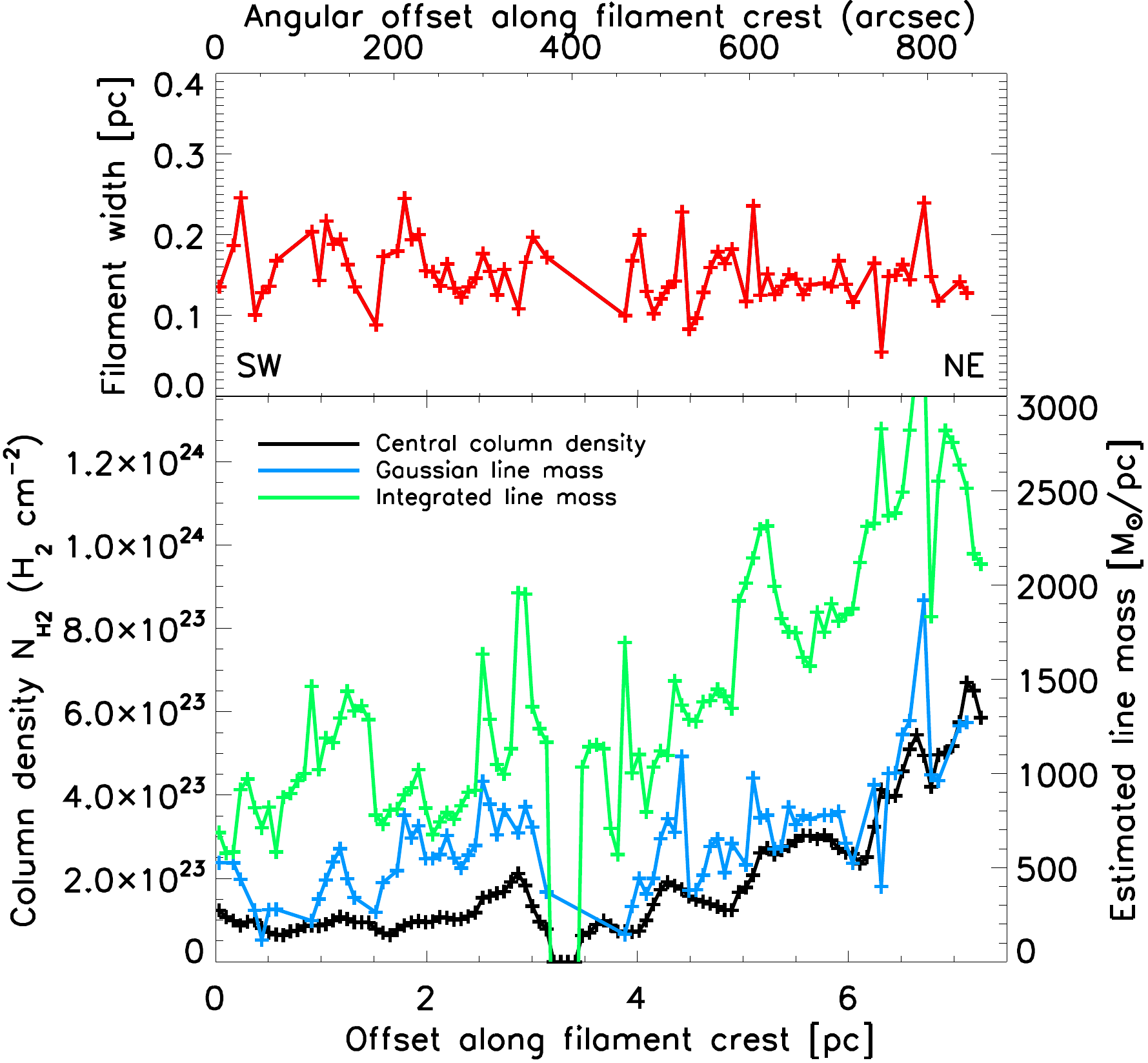}} 
   \caption{
   {\bf(a)} Median radial intensity profile of the NGC~6334 filament (black solid curve) measured perpendicular to, and on the western side of, 
   the filament crest shown as a magenta curve in Fig.~\ref{artemis_maps}b. The yellow error bars visualize the ($\pm 1\sigma $) dispersion of the {\it distribution} of 
   radial intensity profiles observed along the filament crest, while the smaller green error bars represent 
   the standard deviation of the {\it mean} intensity profile (the data points and error bars are spaced by half a beam width). 
   The blue solid curve shows the effective beam profile of the ArT\'eMiS 350 $\mu$m data as measured on Mars, on top of 
   a constant level corresponding to the typical background intensity level observed at large radii. The blue dotted curve shows the best-fit 
   Gaussian ($+$ constant offset) model to the inner part of the observed profile. 
   The red dashed curve shows the best-fit Plummer model convolved with the beam as described in the text.
   {\bf(b)} {\it Upper panel:} Deconvolved Gaussian FWHM width of the NGC~6334 filament as a function of position along the filament crest 
   (from south-west to north-east), with one width measurement per independent ArT\'eMiS beam of $8\arcsec $ (red crosses).
   {\it Lower panel:} Observed column density (black curve and left y-axis), background-subtracted mass per unit length (blue curve and right y-axis) 
   derived from a Gaussian fit (cf. blue dotted curve in {\bf(a)}) to the filament transverse profile, and background-subtracted mass per unit length 
   (green curve and right y-axis) 
   derived from integration of the observed transverse profile (cf. black curve in {\bf(a)}) up to an outer radius of $0.7\, $pc (or $\sim 80\arcsec $) where 
   the background starts to dominate, all plotted as a function of position along the filament crest.
 }
              \label{filament_profile}
\end{figure*}
 
By taking perpendicular cuts at each pixel along the crest, we 
constructed radial intensity profiles for the main filament.
The western part of the resulting median radial intensity profile is displayed in log-log format in Fig.~\ref{filament_profile}a. 
Since at least in projection there appears to be a gap roughly in the middle of the filament crest (cf. Fig.~\ref{artemis_maps}), 
we also divided the filament into two parts, a northern and a southern segment, shown by 
the white and the magenta curve in Fig.~\ref{artemis_add_maps}a, respectively. 
The gap between the two segments may have been created by an HII region visible 
as an H$\alpha$ nebulosity in Fig.~\ref{artemis_add_maps}b. 
Separate radial intensity profiles for the northern and southern segments are shown 
in Fig.~\ref{north_filament_profile} and Fig.~\ref{south_filament_profile}, respectively. 
Due to the presence of the two massive protostellar clumps NGC$\,$6334$\,$I and I(N) \citep[][see also Fig.~\ref{artemis_add_maps}a]{Sandell2000}, 
it is difficult to perform a meaningful radial profile analysis on the eastern side of the northern segment, 
and the corresponding intensity profile has larger error bars (cf. Fig.~\ref{north_filament_profile}a). 

Following \citet{Arzoumanian+2011} and \citet{Palmeirim+2013} 
we fitted 
each radial profile $I (r)$ observed as a 
function of radius $r$ with both a simple Gaussian model: 

\begin{equation}
 I_{\rm G} (r) = I_0 \times {\rm exp}\left[-4\, {\rm ln\,2}\times \left({r/\overline{{\rm FWHM}}}  \right)^2 \right] + I_{Bg},
\end{equation}

\noindent
and a Plummer-like model function of the form:

\begin{equation}
 I_{p} (r) = \,  \frac{I_0}{\left[1+\left({r/R_{\rm flat}}\right)^{2}\right]^{\frac{p-1}{2}}} + I_{Bg},
\end{equation}

\noindent
where $I_0$ is the central peak intensity, $\overline{{\rm FWHM}}$ is the physical FWHM width of the Gaussian model, 
$R_{\rm flat}$ the characteristic radius 
of the flat inner part of the model profile, $p > 1$ is the power-law index of the underlying density profile (see below), 
and $I_{Bg}$ is a constant background level. 
The above functional forms were convolved 
with the approximately Gaussian beam of the ArT\'eMiS data (FWHM $\sim 8\arcsec $) 
prior to comparison with the observed profile. 
The best-fit Gaussian and Plummer-like models for the median radial intensity profile 
observed on the western side of the entire filament 
are shown by the blue dotted and 
red dashed curves in Fig.~\ref{filament_profile}a, respectively. 
Note that only the inner part of the radial profile was fitted with a Gaussian model since 
the observed profile includes an approximately power-law wing which cannot be reproduced 
by a Gaussian curve (cf. Fig.~\ref{filament_profile}a). In practice, a background level was first 
estimated as the intensity level observed at the closest point to the filament's crest for which the logarithmic slope 
of the radial intensity profile $ {\rm d\, ln}\, {I}/{\rm d\, ln}\,r $ became significantly positive. 
This allowed us to obtain a crude estimate of the width of the profile at half power above the background level, 
and the observed profile was then fitted with a Gaussian model over twice this initial width estimate. 
The deconvolved 
diameter of the best-fit Gaussian model is $\overline{{\rm FWHM}} = 0.15 \pm 0.02\, $pc 
and the diameter of the inner plateau in the best-fit Plummer model is $2\, R_{\rm flat} = 0.11 \pm 0.03\, $pc. 
The power-law index of the best-fit Plummer model is $p = 2.2 \pm 0.3$. 
Assuming optically thin dust emission at $350\, \mu$m and using the dust temperature map derived from 
{\it Herschel} data at $36.3\arcsec $ resolution \citep{Russeil+2013,Tige+2016}, 
we also converted the $350\, \mu$m image 
of  Fig.~\ref{artemis_maps}b ($I_{350}$) into an approximate column density image (see Fig.~\ref{artemis_add_maps}a)  
from the simple relation $N_{\rm H_2} = I_{350}/(B_{350}[T_{\rm d}] \kappa_{350} \mu_{\rm H_2} m_{\rm H})$, 
where $B_{350}$ is the Planck function, $T_{\rm d}$ the dust temperature, $ \kappa_{350}$ the dust opacity 
at $\lambda = 350\, \mu$m, and $\mu_{\rm H_2}=2.8$ the mean molecular weight. 
We adopted the same dust opacity law as in our HGBS and HOBYS papers: $\kappa_{\lambda} = 0.1 \times (\lambda/300~\mu \rm m)^{-\beta}$
cm$^{2}$ per g (of gas $+$ dust) with an emissivity index $\beta =2$ \citep[][]{Hildebrand1983,Roy+2014}. 
The y-axis shown on the right of Fig.~\ref{filament_profile}a gives an approximate column density scale 
derived in this way for the median radial profile of the filament 
assuming a uniform temperature $T_{\rm d} = 20\,$K, which corresponds to the median dust temperature
derived from {\it Herschel} data along the crest of the filament.
We also derived and fitted a median radial column density profile for the filament directly using the column density map 
(see Fig.~\ref{coldens_filament_profile} in Appendix~A). 
The results of our radial profile analysis for the whole filament and its two separate segments 
are summarized in Table~\ref{table_comp}, which also provides a comparison 
with similar measurements reported in the recent literature
for four other well-documented filaments. 

\begin{table*}  
\begin{minipage}{1\linewidth}     
\centering
\caption{Derived properties of the NGC~6334 filament and comparison with other well-documented filaments}
\label{table_comp}
\vspace*{-0.25cm}
\begin{tabular}{c|c|c|c|c|c|c|c|c}   
\hline\hline   
Filament  & $\langle M_{\rm line} \rangle $ ~\tablefootmark{a}   &  $\langle N^0_{\rm H_{2}} \rangle $ ~\tablefootmark{b}   & $N^{\rm bg}_{\rm H_{2}}  ~~\tablefootmark{c} $ &   $p$ ~~\tablefootmark{d}    &   $R_{\rm flat}$ ~\tablefootmark{e}   & Width, $W$ ~\tablefootmark{f}  & Length 	& Refs \\
	        & ($M_\odot $/pc)      &   (${\rm cm}^{-2} $)     &    (${\rm cm}^{-2} $)             &           &   (pc)     & (pc)    & (pc)  &          \\      
\hline
NGC~6334 north$+$south & 800--1300 &  1--2$\times 10^{23} $  & $1.8 \times 10^{22} $ &   $2.2 \pm 0.3$   &   $0.05\pm 0.01$        & $0.15 \pm 0.03 $ & $> 7$     & 1 \\
(western side) 			 &  		       &    				      &  				  &      		       &          			  &  				&                 &    \\
NGC~6334 north$+$south & (900--1300) &  (1--2$\times 10^{23} $)  & (2--4$\times 10^{22} $) &   ($1.9 \pm 0.4$)   &   ($0.05\pm 0.02$)        & ($0.19 \pm 0.03$)  & $> 7$     & 1 \\
(eastern side) 			 &  		       &    				      &  				  &      		       &          			  &  				&                 &    \\
NGC~6334 north                 &      1600      &   $2.5 \times 10^{23} $  & $2.1 \times 10^{22} $ &   $2.4 \pm 0.3$   &   $0.06\pm 0.02$        & $0.15 \pm 0.03 $ & $> 3.5$  & 1 \\
(western side) 			 &  		       &    				      &  				  &      		       &          			  &  				&                 &    \\
NGC~6334 north\tablefootmark{g}  & (800--1600)  &   (1.5--2.5$\times 10^{23}) $  & ($1.1 \times 10^{23} $) &   --   &   --        & ($0.20 \pm 0.03 $) & $> 3.5$  & 1 \\
(eastern side) 			 &  		       &    				      &  				  &      		       &          			  &  				&                 &    \\
NGC~6334 south                &  500--600  &  0.7--1$\times 10^{23} $  & $1.2 \times 10^{22} $ &  ($2.3 \pm 0.3$)  &  ($0.07\pm 0.02$)   & $0.16 \pm 0.04 $ & $\sim 3 $ & 1 \\
(western side) 			 &  		       &    				      &  				  &      		       &          			  &  				&                  &    \\
NGC~6334 south                &  700--2000  &   0.9--1$\times 10^{23} $   &  $1.2 \times 10^{22} $           &  $1.8 \pm 0.3$    &    $0.09\pm 0.02$      & $0.16 \pm 0.04 $  & $\sim 3 $ & 1 \\
(eastern side) 			 &  		       &    				      &  				  &      		       &          			  &  				&                  &    \\
\hline
Vela~C\tablefootmark{h} 				 & 320--400    &  $ 8.6 \times 10^{22} $  & $3.6  \times 10^{21} $ &   $2.7 \pm 0.2$   &   $0.05\pm 0.02$        & $0.12 \pm 0.02 $ &     $4$     & 2 \\
Serpens South	                   &     290          &  $ 6.4 \times 10^{22} $  & $3.7  \times 10^{21} $ &   $2.0 \pm 0.3$   &   $0.03\pm 0.01$        & $0.10 \pm 0.05 $ &     $2$      & 2, 3 \\
Taurus B211/B213	          &       50          &  $ 1.5 \times 10^{22} $  & $0.7  \times 10^{21} $ &   $2.0 \pm 0.3$   &   $0.03\pm 0.02$        & $0.09 \pm 0.02 $ &     $> 5$   & 4  \\
Musca				 &      20           &  $ 4.2 \times 10^{21} $  & $0.8  \times 10^{21} $ &   $2.2 \pm 0.3$   &   $0.08$        & $0.14 \pm 0.03 $ &     $10$    & 5, 6 \\
                           \hline  
                  \end{tabular}
\tablefoot{
Values given in parentheses are more uncertain due to, e.g., large error bars in the corresponding filament profiles, and should be understood as being only indicative.\\
 \tablefoottext{a}{Average mass per unit length of the equivalent cylindrical filament derived from one-sided integration of the observed radial column density profile after background subtraction.   
The actual mass per unit length of each filament segment corresponds to the mean of the eastern-side and western-side values (not explicitly given here).
The outer radius of integration was 0.7~pc on the western side, 0.3~pc on the eastern side of the northern segment, and 1.4~pc on the eastern side of the southern segment, respectively, corresponding  
 to the radius where the background starts to dominate (see  
 Figs.~\ref{north_filament_profile}--\ref{coldens_filament_profile}). }\\
 \tablefoottext{b}{Average value of the central column density derived along the filament crest after background subtraction. 
Typical uncertainties are a factor $\sim \, $1.5--2 for values $< 10^{23} {\rm cm}^{-2} $ and a factor $\sim \, $2--3 for values $> 10^{23} {\rm cm}^{-2} $, dominated by 
uncertainties in the dust opacity and in the distribution of dust temperature along the line of sight \citep{Roy+2014}.}\\
 \tablefoottext{c}{Background column density. This is estimated as the column density observed at the closest point to the filament's crest for which the logarithmic slope of the radial column density profile 
$ {\rm d\, ln}\, {N_{\rm H_2}}/{\rm d\, ln}\,r $ becomes positive.}\\ 
\tablefoottext{d}{Power-law index of the best-fit Plummer model [see Eq.~(2)].}\\
\tablefoottext{e}{Radius of the flat inner plateau in the best-fit Plummer model  [see Eq.~(2)].}\\
\tablefoottext{f}{Deconvolved FWHM width from a Gaussian fit to the inner part of the filament profile.}\\
\tablefoottext{g}{The eastern side of the radial column density profile of the northern filament is poorly constrained due to confusion with the two massive protostellar clumps 
NGC$\,$6334$\,$I and I(N) (cf. Fig.~\ref{artemis_add_maps}a); no meaningful Plummer fit is possible.}\\
\tablefoottext{h} {According to \citet{Minier+2013}, the Vela~C filament is not a simple linear structure or ``ridge'', but is part of a more complex ring-like structure at least partly shaped
by ionization associated with the RCW~36 HII region.}
\tablebib{(1)~this paper;
(2) \citet{Hill+2012}; (3) \citet{Konyves+2015}; (4) \citet{Palmeirim+2013};
(5) \citet{Cox+2016}; (6) \citet{Kainulainen+2016}.
}
}
 \end{minipage}
\end{table*}

We stress that the presence of cores along the filament has virtually no influence on 
the results reported in Table~\ref{table_comp}. 
First, as already mentioned the portion of the filament selected here contains only one candidate 
starless MDC at the northern end \citep[cf.][]{Tige+2016}, and the width estimates 
are unchanged when the immediate vicinity of this object is excluded from the analysis 
(see also Fig.~\ref{filament_profile}b). Second, low-mass prestellar cores typically contribute only a small fraction ($\simlt 15\% $)
of the mass of dense filaments \citep[e.g.][]{Konyves+2015}. 
Third, we performed the same radial profile analysis on a source-subtracted image generated 
by \textsl{getsources} \citep{Menshchikov2012} and obtained very similar results.

One advantage of the Plummer-like functional form in Eq.~(2) is that, when applied to a filament column density profile 
($I_0$ becoming $N_{\rm H_2, 0}$, the central column density), it directly informs 
about the underlying volume density profile, which takes a similar form, n$_{p}(r) = \frac{{\rm n}_{\rm H_2, 0}}{\left[1+\left({r/R_{\rm flat}}\right)^{2}\right]^{p/2}}$, 
where n$_{\rm H_2, 0}$ is the central volume density of the filament. 
The latter is related to the projected central column density $N_{\rm H_2, 0}$ by the simple relation, 
n$_{\rm H_2, 0} = N_{\rm H_2, 0}/(A_p\, R_{\rm flat}) $, where $A_p = \frac{1}{\cos\,i} \times B\left(\frac{1}{2},\frac{p-1}{2}\right) $ is a constant factor 
taking into account the filament's inclination angle to the plane of the sky, and $ B$ is the Euler beta function \citep[cf.][]{Palmeirim+2013}.
Here, assuming $i = 0\degr $, we estimate the mean central density 
to be n$_{\rm H_2, 0} \sim 2.2 \times 10^5\, {\rm cm}^{-3} $,  $\sim 5 \times 10^5\, {\rm cm}^{-3} $, 
and $\sim 1.5 \times 10^5\, {\rm cm}^{-3} $ in the entire filament, the northern segment, 
and the southern segment, respectively.

\section{Discussion and conclusions}
\label{dis}

Our ArT\'eMiS mapping study confirms that the main filament in NGC~6334 is a very dense, massive linear structure 
with $M_{\rm line}$ ranging from $\sim 500\, M_\odot$/pc to $\sim 2000\, M_\odot$/pc over nearly $10\,$pc, 
and demonstrates for the first time that its inner width remains as narrow as $W \sim 0.15\pm 0.04\,$pc all along 
the filament length (see Fig.~\ref{filament_profile}b), within a factor of $< 2$ of the characteristic $0.1\,$pc value found by \citet{Arzoumanian+2011} 
for lower-density nearby filaments in the Gould Belt. 

While the NGC~6334 filament is highly supercritical, and of the same order of magnitude in line mass 
as high-mass star-forming 
ridges such as DR21 \citep{Schneider+2010, Hennemann+2012}, it is remarkably simple 
and apparently consists of only a single, narrow linear structure. 
In contrast, a massive ridge is typically resolved into a closely packed network of sub-filaments and 
``massive dense cores'' (MDCs) \citep{Motte+2016}. 
This is at variance with the NGC~6334 filament which exhibits a surprisingly low level of fragmentation. 
The maximum relative column density fluctuations observed along its long axis (cf. black curve in Fig.~\ref{filament_profile}b) 
are only marginally nonlinear ($\delta N_{\rm H_2}/\langle N_{\rm H_{2}} \rangle  \approx 1$), 
while for instance most of the supercritical low-mass filaments analyzed by \citet{Roy+2015} 
have stronger fluctuations (with $\delta N_{\rm H_2}/\langle N_{\rm H_{2}} \rangle $ up to $\sim \,$2--5). 
Most importantly, the NGC~6334 filament harbors no MDC, 
except perhaps at its two extremities \citep[][see Fig.~\ref{artemis_add_maps}a]{Tige+2016}. 
It is therefore unclear whether the filament will form high-mass stars 
or not. 
On the one hand, the lack of MDCs suggests that the filament may not  form any massive stars in the near future. 
On the other hand, the presence of a compact HII region (radio source C from \citealp{Rodriguez+1982}) 
at the north-east end of the southern 
part of the filament, near the gap between the two filament segments (see Fig.~\ref{artemis_add_maps}a), 
suggests that it may have already formed massive stars in the past. 
Based 
on observations in the HCO$^+$(3--2) and  H$^{13}$CO$^+$(1--0) lines with APEX and MOPRA, 
\citet{Zernickel+2013} showed that the 
filament is coherent in velocity and 
found a velocity gradient of $\sim 2\, {\rm km\, s^{-1}/pc} $ from both ends of the filament 
toward its center. They 
proposed that 
the whole filament is 
in a state of global collapse along its long axis toward its center (estimated to be close 
to the gap between the two segments in Fig.~\ref{artemis_add_maps}a). 
This proposal is qualitatively consistent with the identification of candidate MDCs at the two ends 
of the filament \citep[][]{Tige+2016}, and with the theoretical expectation that the longitudinal collapse 
of a finite filament is end-dominated due to maximal gravitational acceleration at the edges
\citep[e.g.][]{Burkert+2004,Clarke+2015}. It is difficult, however, to explain the presence of HII regions 
-- significantly more evolved than MDCs -- near the central gap 
in this picture, unless these HII regions did not form in the filament but in the vicinity 
and disrupted the central part of the filament. 

The low level of fragmentation poses a challenge to theoretical models since supercritical filaments are supposed 
to contract radially and fragment along their length in only about one free-fall time or $\sim\, $4.5--8\,$\times 10^4\,$yr in the present 
case \citep[e.g.][]{InutsukaMiyama1997}. One possibility is that the NGC~6334 filament is observed at a very early 
stage after its formation by large-scale compression. 
Another possibility is that the filament is ``dynamically'' supported against rapid radial contraction and longitudinal 
fragmentation by accretion-driven MHD waves \citep[cf.][]{HennebelleAndre2013}. 
The average one-dimensional velocity dispersion $\sigma _{\rm 1D}$ estimated from the $40\arcsec $ resolution N$_2$H$^+$(1--0) observations 
of the MALT90 survey with the MOPRA telescope \citep[][]{Jackson+2013} is $ \sim 1.1\,$km/s 
in the northern part of the filament and $ \sim 0.7\,$km/s in the southern segment. 
Compared to the sound speed $c_s \sim  0.3\, $km/s given an estimated gas temperature $T \sim \,$20--25$\,$K, 
this velocity dispersion is supersonic by a factor $\sim\, $2--4, implying that there may be significant velocity substructure 
(such as the presence of several sonic velocity components -- cf. \citealp{Hacar+2013}) 
in the filament. 
Ignoring any static magnetic field, the virial mass per unit length, $M_{\rm line, vir} = 2\, \sigma _{\rm 1D}^2/G $ \citep[cf.][]{Fiege2000}, 
is thus $\sim 560\, M_\odot $/pc and $\sim 220\, M_\odot $/pc in the northern and southern segments, respectively, 
which is consistent with the filament being within a factor of $\sim 2$ of virial balance. 
A static magnetic field can easily modify $M_{\rm line, vir}$ by a factor of $\sim 2$ \citep[cf.][]{Fiege2000}, 
and a significant static field component perpendicular to the long axis of the filament would help 
to resist collapse and fragmentation along the filament. 
Higher-resolution observations in molecular line tracers of dense gas would be needed to investigate 
whether the NGC~6334 filament contains a bundle of intertwined velocity-coherent fibers similar to 
the fibers identified by \citet{Hacar+2013} in the low-mass B211--3 filament in Taurus. 
The detection of such braid-like velocity substructure may provide indirect evidence of the presence of internal MHD waves.

In any case, and regardless of whether the NGC~6334 filament will form massive stars or not,  
our ArT\'eMiS result that the filament inner width is within a factor of 2 of $0.1\,$pc has interesting implications. 
Our NGC~6334 study is clearly insufficient to prove that interstellar filaments 
have a truly universal inner width, but it shows 
that the finding obtained with {\it Herschel} in nearby clouds is not limited to filaments in low-mass star forming regions.  
It is quite remarkable that the NGC~6334 filament has almost the same inner width as 
the faint subcritical filaments in Polaris \citep[cf.][]{Menshchikov+2010,Arzoumanian+2011}, 
the marginally supercritical filaments in Musca and Taurus \citep[][]{Cox+2016, Palmeirim+2013}, 
or the lower-mass supercritical filaments in Serpens South and Vela~C \citep[][]{Hill+2012}, 
despite being three orders of magnitude, two orders of magnitude, and at least a factor of $\sim \, $3  
denser and more massive than these filaments, respectively
(see Table~\ref{table_comp}). 
While not all of these filaments may have necessarily formed in the same way, this suggests that a common physical 
mechanism is responsible for setting the filament width at the formation stage 
and that the subsequent evolution of dense filaments -- through, e.g., accretion of background cloud material \citep[cf.][]{Heitsch2013, HennebelleAndre2013} -- is such 
that the inner width remains at least approximately conserved with time. 
A promising mechanism for creating dense filaments, which may be quite generic especially in massive star-forming 
complexes, 
is based on multiple episodes of large-scale supersonic compression due to interaction of expanding bubbles \citep{Inutsuka+2015}.
With about 7 bubble-like HII regions per square degree (\citealp{Russeil+2013}, see also Fig.~\ref{artemis_add_maps}), 
there is ample opportunity for this mechanism to operate 
in NGC~6334.  
More specifically, at least in projection, the NGC~6334 filament appears to be part of an arc-like structure 
centered on the HII region Gum~63  (see Fig.~\ref{artemis_add_maps}a), suggesting the filament may partly 
result from the expansion of the associated bubble.
Interestingly, 
the background column density is 
one order of magnitude higher 
for 
the NGC~6334 filament 
than for the other filaments of Table~\ref{table_comp}, 
which is suggestive of a significantly stronger compression.
Further observational studies will be needed to investigate the structure and environment of a larger number of filaments  
in massive star forming regions, and will determine whether the characteristics of the NGC~6334 filament are generic or not.
More theoretical work is also needed to better understand the physics controlling the width of interstellar filaments. 

\begin{acknowledgements}
We would like to thank ESO and the APEX staff in Chile for their support 
of the ArT\'eMiS project. We acknowledge financial support from 
the French National Research Agency (Grants ANR--05--BLAN-0215 \& ANR--11--BS56--0010, 
and LabEx FOCUS ANR--11--LABX-0013). 
Part of this work was also supported by the European Research Council 
under the European Union's Seventh Framework Programme 
(ERC Advanced Grant Agreement no. 291294 --  `ORISTARS').
This research has made use of data from the {\it Herschel} HOBYS project (http://hobys-herschel.cea.fr). 
HOBYS is a {\it Herschel} Key Project jointly carried out by SPIRE Specialist Astronomy Group 3 (SAG3), scientists 
of the LAM laboratory in Marseille, and scientists of the  {\it Herschel} Science Center (HSC).
\end{acknowledgements}

%
%

\bibliographystyle{aa}
\bibliography{ngc6334_artemis,ref,ref_fil}


\newpage

\begin{appendix} 
\vspace*{-0.35cm}
\section{Additional radial profiles}\label{sec:appendix}

\begin{figure*}[!h]
\vspace*{1.0cm}
      \centering
    \resizebox{9.cm}{!}{      
\includegraphics[angle=0]{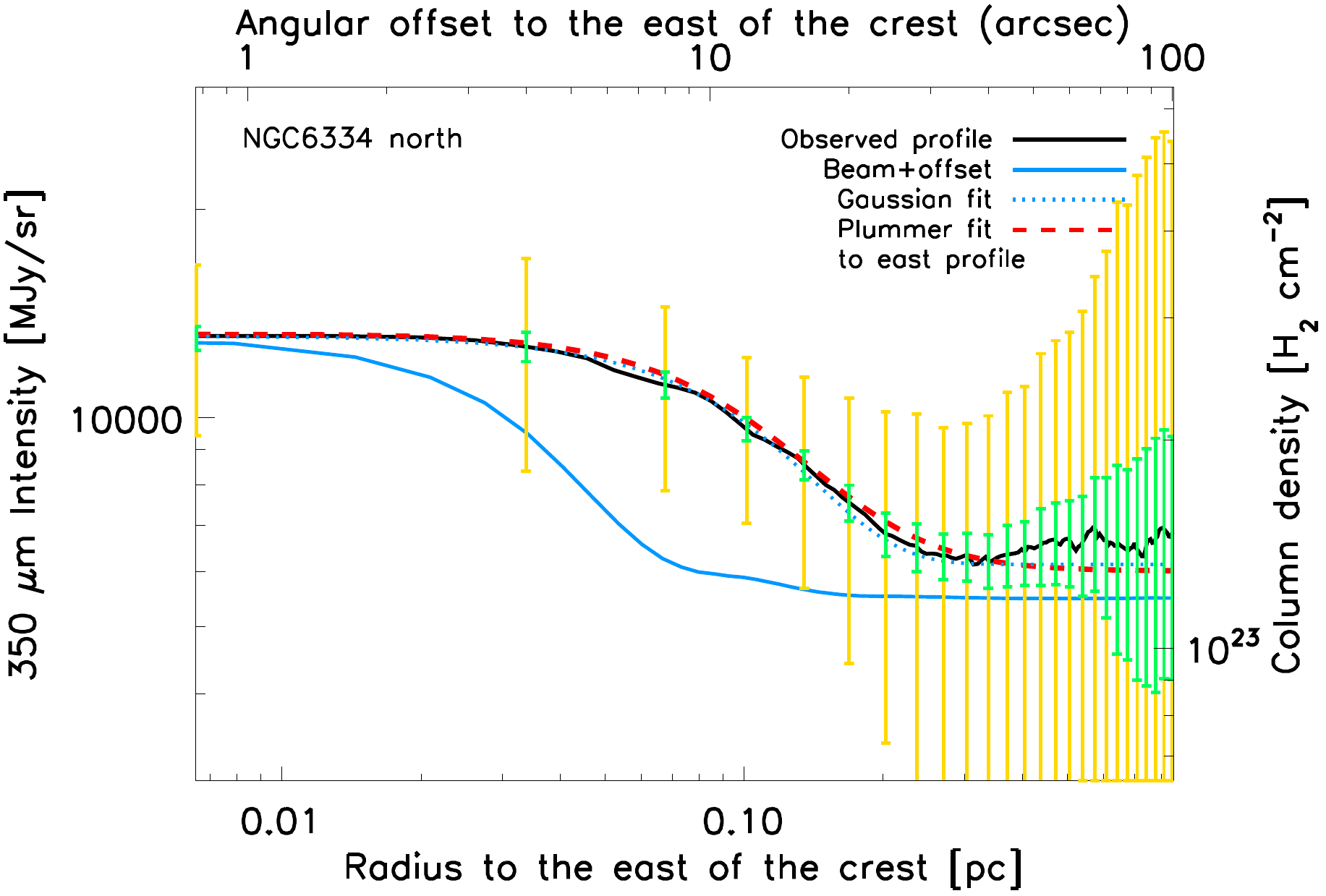}}
       \hspace{1mm}
  \resizebox{9.cm}{!}{
\includegraphics[angle=0]{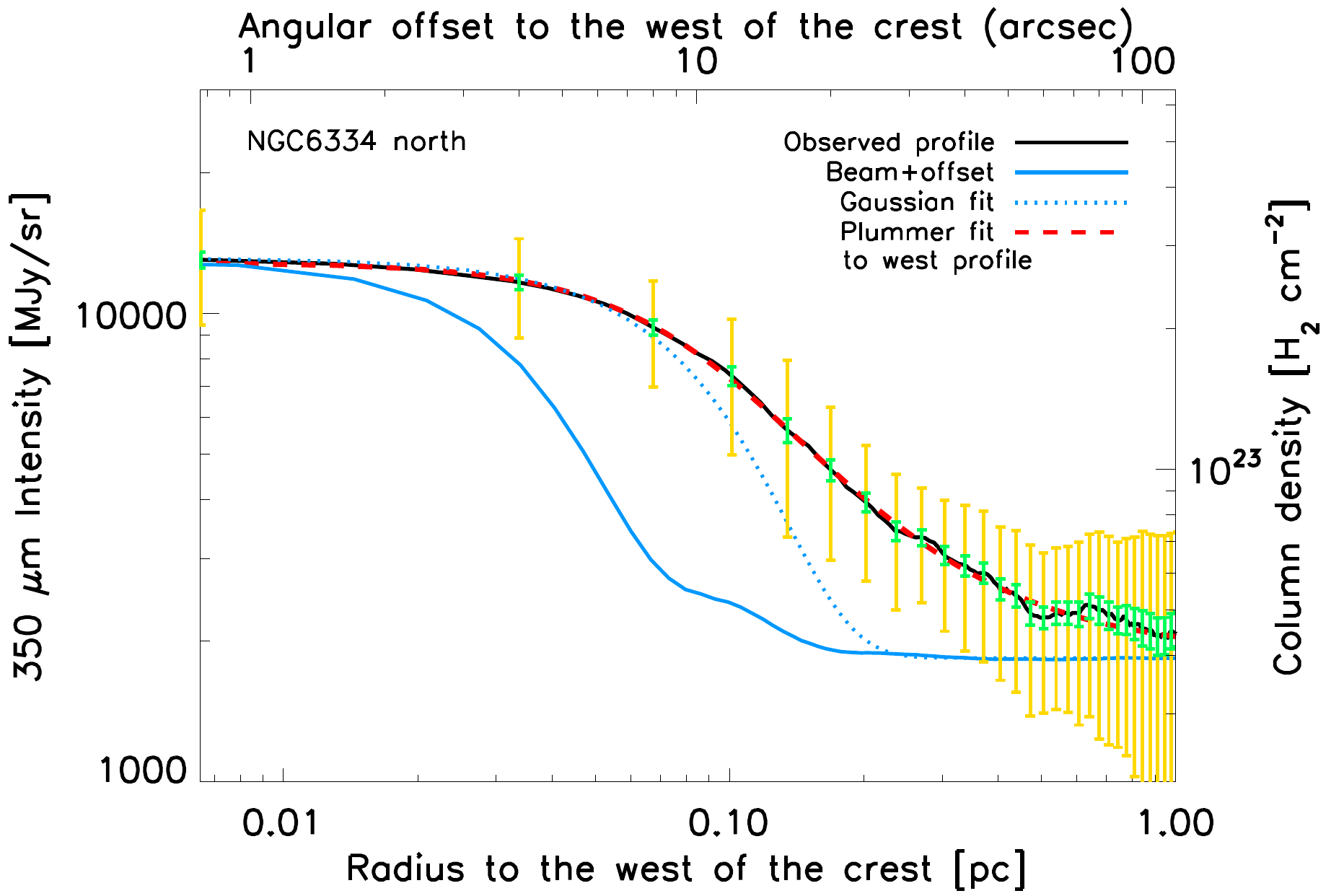}}
   \caption{
   {\bf(a)} Median radial intensity profile of the northern part of the NGC~6334 filament (black solid curve) measured in  
   the combined ArT\'eMiS $+$ SPIRE 350 $\mu$m image (Fig.~\ref{artemis_maps}b) 
   perpendicular to, and on the eastern side of, 
   the filament crest shown as a white curve in Fig.~\ref{artemis_add_maps}a. The yellow and green error bars are as in Fig.~\ref{filament_profile}a. 
    The blue solid curve shows the effective beam profile of the ArT\'eMiS 350 $\mu$m data as measured on Mars, on top of 
   a constant level corresponding to the typical background intensity level observed at large radii. The blue dotted curve shows the best-fit 
   Gaussian ($+$ constant offset) model to the inner part of the observed profile. 
   The red dashed curve shows the best-fit Plummer model convolved with the beam [cf. Sect.~\ref{obs_ana} and Eq.~(2)]. 
   {\bf(b)} 
   Same as in {\bf(a)} but for the median radial intensity profile of the northern part of the 
   filament measured on the western side of 
    the filament crest shown as a white curve in Fig.~\ref{artemis_add_maps}a.
 }
              \label{north_filament_profile}

\vspace*{2.0cm}

      \centering
    \resizebox{9.cm}{!}{      
\includegraphics[angle=0]{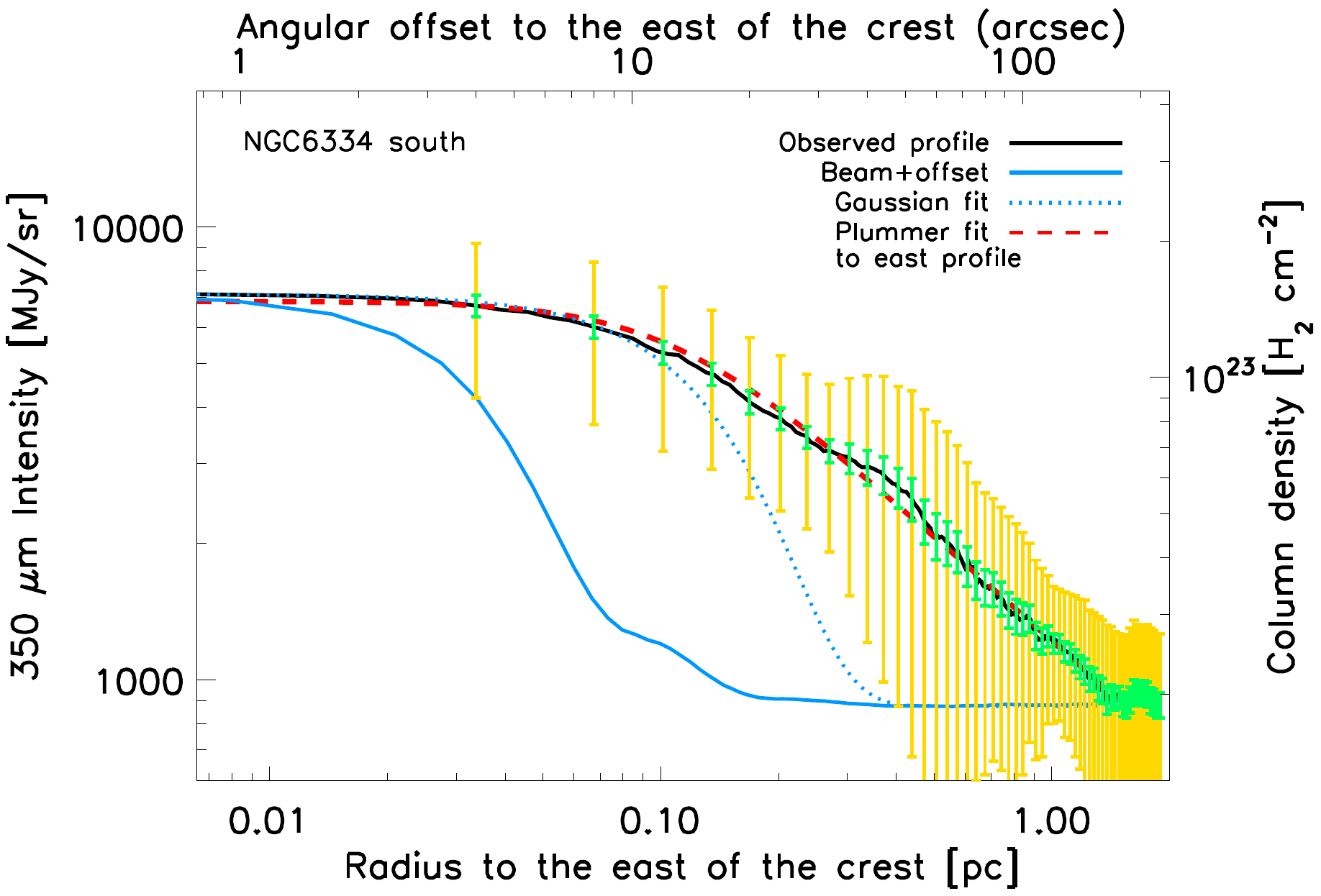}}
       \hspace{1mm}
  \resizebox{9.cm}{!}{
 \includegraphics[angle=0]{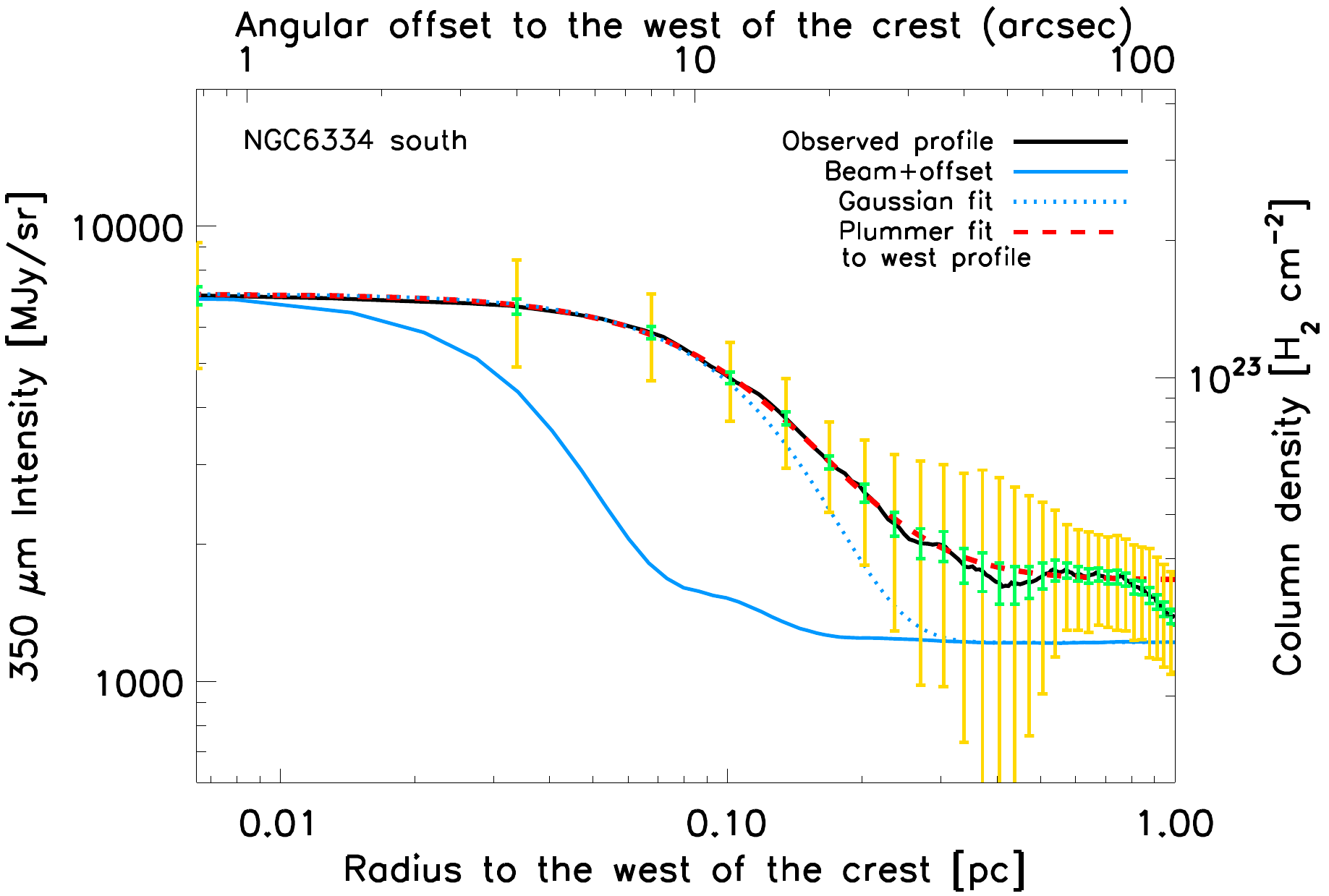}} 
   \caption{
   {\bf(a)} Median radial intensity profile of the southern part of the NGC~6334 filament (black solid curve) measured in  
   the combined ArT\'eMiS $+$ SPIRE 350 $\mu$m image (Fig.~\ref{artemis_maps}b)    
   perpendicular to, and on the eastern side of, 
   the filament crest shown as a magenta curve in Fig.~\ref{artemis_add_maps}a. The yellow and green error bars are as in Fig.~\ref{filament_profile}a.
    The blue solid curve shows the effective beam profile of the ArT\'eMiS 350 $\mu$m data as measured on Mars, on top of 
   a constant level corresponding to the typical background intensity level observed at large radii. The blue dotted curve shows the best-fit 
   Gaussian ($+$ constant offset) model to the inner part of the observed profile. 
   The red dashed curve shows the best-fit Plummer model convolved with the beam [cf. Sect.~\ref{obs_ana} and Eq.~(2)].
   {\bf(b)} 
   Same as in {\bf(a)} but for the median radial intensity profile of the southern part of the 
   filament measured on the western side of 
    the filament crest shown as a magenta curve in Fig.~\ref{artemis_add_maps}a.
 }
              \label{south_filament_profile}
\end{figure*}

\begin{figure*}[!h]
\vspace*{1.0cm}
      \centering
    \resizebox{8.5cm}{!}{      
\includegraphics[angle=0]{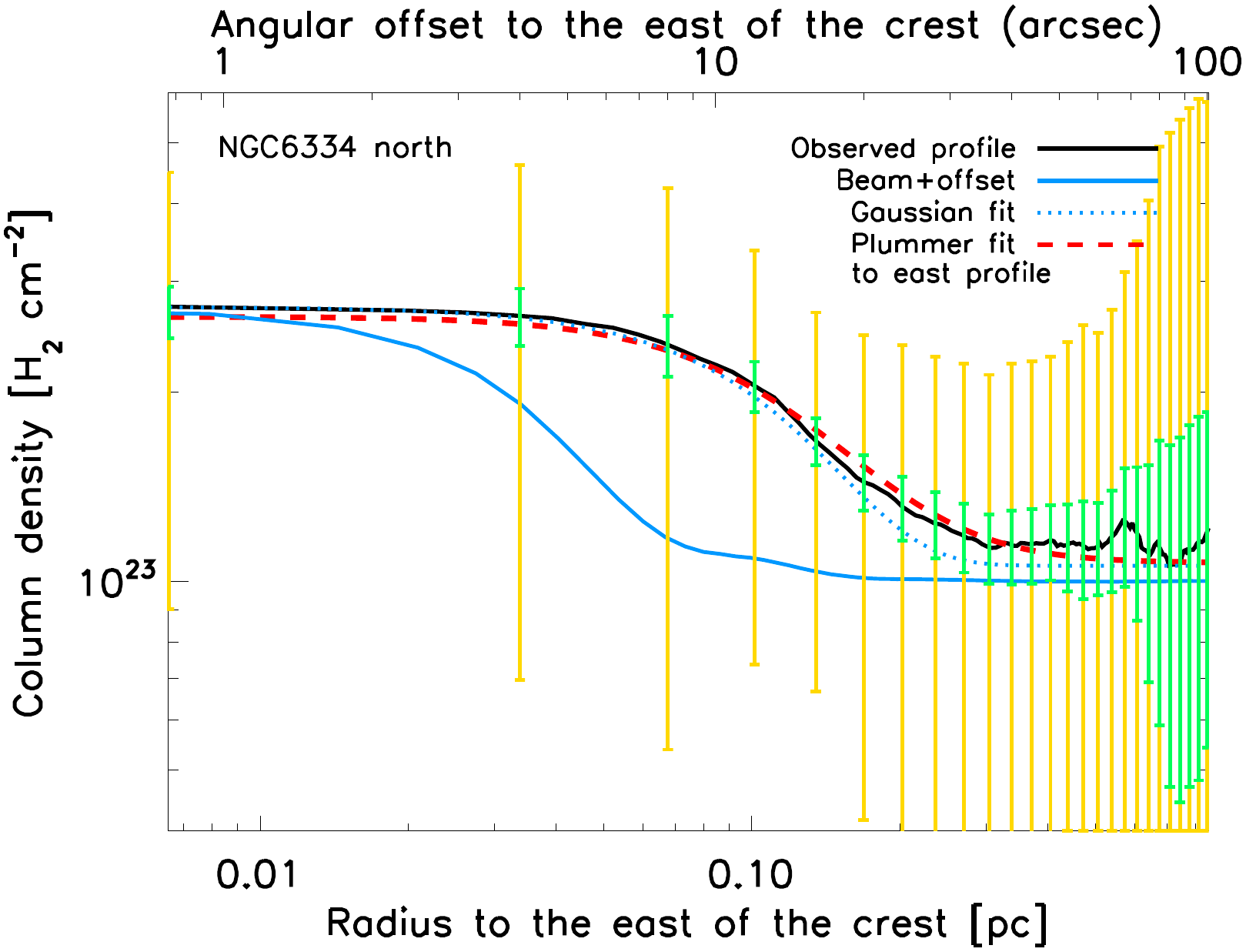}}
       \hspace{5mm}
  \resizebox{8.5cm}{!}{
\includegraphics[angle=0]{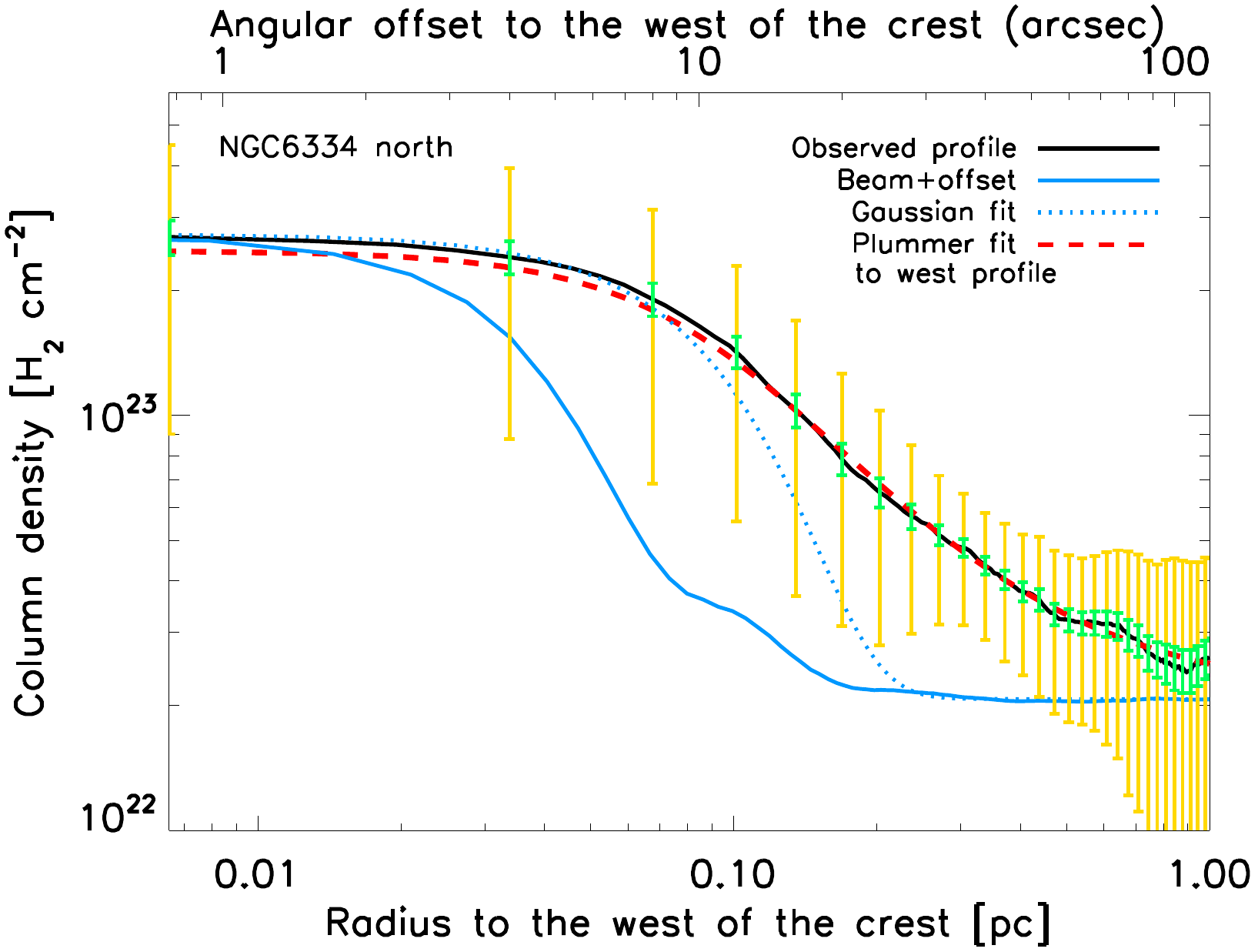}} 
   \caption{
   {\bf(a)} Same as Fig.~\ref{north_filament_profile}a but for the median radial column density profile measured on the eastern side of the northern part of the  
   NGC~6334 filament in the approximate  column density map shown in Fig.~\ref{artemis_add_maps}a.
   {\bf(b)} Same as Fig.~\ref{north_filament_profile}b but for the median radial column density profile measured on the western side of the northern part of the  
   filament in the column density map shown in Fig.~\ref{artemis_add_maps}a.
 }
              \label{coldens_filament_profile}
\end{figure*}

\end{appendix}

\end{document}